\documentclass[reprint,prl,twocolumn]{revtex4-1}

\usepackage{amsmath,amssymb}
\usepackage{graphicx}
\usepackage{microtype}
\usepackage{nicefrac}
\usepackage{physics}
\usepackage{mathrsfs}
\usepackage{slashed}

\usepackage[english]{babel}
\usepackage[applemac]{inputenc}
\usepackage[T1]{fontenc}

\newcommand{\beq}{\begin{equation}}
\newcommand{\eeq}{\end{equation}}
\newcommand{\beqa}{\begin{eqnarray}}
\newcommand{\eeqa}{\end{eqnarray}}

\usepackage{soul}
\usepackage[normalem]{ulem}

\usepackage{enumitem}
\setlist[description]{leftmargin=*}

\usepackage[usenames,dvipsnames]{color}
\usepackage{hyperref}
\hypersetup{colorlinks=true,linkcolor=blue,citecolor=blue,filecolor=green,urlcolor=blue}

\let\oldparagraph\paragraph
\renewcommand{\paragraph}[1]{\oldparagraph{\textbf{#1}}}

\begin{document}

\title{Diagrammatic Monte Carlo for electronic correlation in molecules: \\ high-order many-body perturbation theory with low scaling}

\author{G.~Bighin$^{1,2}$, Q.~P.~Ho$^{2,3}$, M.~Lemeshko$^{2}$, T.~V.~Tscherbul$^{4}$}

\affiliation{$^1$Institut f\"ur Theoretische Physik, Universit\"at Heidelberg, Philosophenweg 19, 69120 Heidelberg, Germany \\
$^{2}$IST Austria (Institute of Science and Technology Austria), Am Campus 1, 3400 Klosterneuburg, Austria\\
$^{3}$Department of Mathematics, Hong Kong University of Science and Technology, Clear Water Bay, Hong Kong \\
$^{4}$Department of Physics, University of Nevada, Reno, NV, 89557, USA}

\date{\today}

\begin{abstract}

We present a low-scaling diagrammatic Monte Carlo approach to molecular correlation energies. Using combinatorial graph theory to encode many-body Hugenholtz diagrams, we sample the M{\o}ller-Plesset (MP$n$) perturbation series, obtaining accurate correlation energies up to $n=5$, with quadratic scaling in the number of basis functions. Our technique reduces the computational complexity of the molecular many-fermion correlation problem, opening up the possibility of low-scaling, accurate stochastic computations for a wide class of many-body systems described by Hugenholtz diagrams.
 
\end{abstract}

\maketitle

The many-electron correlation energy, defined as the difference between the true energy of a many-electron system and that obtained in the Hartree-Fock (HF) approximation \cite{Szabo:1996,Bartlett:07,Shavitt:09,Cremer:11,Hirata:14,Tew:07}, plays a central role in the theoretical description of a wide array of phenomena in chemistry, physics, and material science  \cite{Szabo:1996,Martin:06,Tew:07}  ranging from dispersion interactions responsible for protein folding \cite{He:09} and photoisomerization of retinal, the first step in vision \cite{Gozem:12,Liu:16} to strongly correlated many-electron states in transition-metal compounds \cite{Chan:11}, and  high-temperature superconductors \cite{Mueck:15}.  
The development of highly efficient computational methods for calculating the many-electron correlation energy is thus an ultimate goal of modern electronic structure theory \cite{Bartlett:07,Shavitt:09,Martin:06}.

Many-body perturbation theory (MBPT) \cite{Szabo:1996,Hirata:14} and coupled-cluster (CC) theory \cite{Bartlett:07} are the two primary methods for treating the effects of dynamic correlation, where a single HF state provides a qualitatively correct zeroth-order approximation to the electronic wavefunction \cite{Bartlett:07}.
Both MBPT and CC methods  have been very successful in predicting the correlation energies of small and medium-size molecules \cite{Bartlett:07,Hirata:14}, whereas second-order M{\o}ller-Plesset perturbation theory (MP2)
\cite{Szabo:1996,Bartlett:07,Shavitt:09} is the method of choice for calculating electron correlation effects in large systems involving thousands of atoms \cite{Hirata:14,Cremer:11}. Despite the well-documented shortcomings of higher-order MP$n$ methods ($n\gtrsim 3$) such as non-monotonous convergence \cite{Cremer:11}, with proper choice of orbitals and regularization, such methods can provide chemically accurate results not only for molecular energy differences, but also for reaction barrier heights and intermolecular interactions \cite{Bertels:19,Hesselmann:19}. 

However, widespread application of MP$n$ and CC methods to larger molecules is limited by the steep scaling of the computational cost ($N^{n+3}$ in the case of MP$n$) with the number of spin-orbitals $N$ \cite{Cremer:11,Hirata:14,Ratcliff:17}.  This problem has motivated the development of ingenious low-scaling methods \cite{Izmaylov:08,Pulay:83,Ayala:01,Werner:03,Izmaylov:08,Neuhauser:13}. Among those, several promising Monte Carlo (MC) techniques rely on stochastic sampling of configuration-interaction (CI) \cite{Thom:07,Booth:09} and CC \cite{Scott:19} expansions in imaginary time or performing real-space  MC integration to obtain MP$n$ energies   \cite{Willow:12,Willow:14,Hirata:14,Li:19,Doran:21}. 

In this Letter, we introduce {a novel} stochastic approach to the many-electron correlation problem in molecules based on the powerful Diagrammatic Monte Carlo (DiagMC) methodology \cite{Prokofev:1998,VanHoucke:2010kya,Gull:11}, which uses direct sampling of the entire diagrammatic series for the many-electron correlation energy to obtain numerically results free of systematic bias.
Originally developed in the context of quantum impurity problems \cite{Prokofev:1998,Gull:11}, DiagMC  has been applied with great success to a wide range of problems in quantum many-body physics, including exotic impurities with internal degrees of freedom \cite{Bighin:2017wx,Bighin:18,Li:2019uv,Li:2020tl}, correlated lattice fermions \cite{Kozik:2010fl}, unitary Fermi gases \cite{VanHoucke:2012ica}, and non-equilibrium quantum dynamics \cite{Cohen:2015de}.  
Recent applications of the DiagMC approach have provided numerically exact correlation energies of the homogeneous electron gas \cite{Chen:2019jp} and of an infinite chain of hydrogen atoms \cite{Motta:17}. Thus far, however,  {save for a very recent application to molecular quantum impurity problems at finite temperature \cite{Li:20}}, DiagMC has not been applied to calculate molecular correlation energies, likely due to the topological complexity of the underlying Hugenholtz diagrams.

Here, we overcome this problem by using combinatorial graph theory to encode Hugenholtz diagrams into adjacency matrices, a technique recently developed in nuclear physics \cite{Tichai:2017,Arthuis:2019}. This allows us to design general and efficient updates for sampling the diagrammatic expansions of MBPT using the Metropolis algorithm.
Unlike full configuration interaction MC \cite{Booth:09}, stochastic MP$n$ theory in real space \cite{Willow:12,Willow:14,Hirata:14,Li:19}, or DiagMC for molecular quantum impurities \cite{Li:20}, our DiagMC/MP$n$ method evaluates the correlation energy directly based on a random walk in the space of Hugenholtz diagrams, rather than that of Slater determinants or in real space.

We apply our approach to calculate the correlation energies of small molecules, obtaining accurate MP$n$ results up to $n=4$ with low $O(N^2)$ scaling, opening up the possibility of accessing heretofore unexplored regimes in the upper right corner of the Pople diagram \cite{Pople:65,Karplus:90} -- i.e.,  computing accurate dynamical correlation energies for much larger systems than was previously possible.
Because our methodology only relies on graph theory, it can be easily extended beyond electronic structure theory to include the diagrammatic expansions that occur in, e.g., 
 vibrational spectroscopy  \cite{Hermes:13,Hermes:14,Hirata:15}, crystal phonon perturbation theory \cite{Cowley:63,Goldman:68,Koehler:69}, and nuclear physics \cite{Tichai:2017,Arthuis:2019}.

\paragraph{MPn theory and matrix encoding of Hugenholtz diagrams.}

In MP$n$ theory  \cite{Hirata:14,Shavitt:09,Cremer:11}, the non-relativistic electronic Hamiltonian $\hat{H}=\sum \varepsilon_i \hat{c}_i^\dag \hat{c}_i + \frac{1}{2}\sum_{ijkl}  \langle ij || kl \rangle \hat{c}^\dag_i \hat{c}_j^\dag \hat{c}_l \hat{c}_k$ is partitioned into the mean-field reference Hamiltonian $\hat{H}_0=\sum_i \varepsilon_i \hat{c}_i^\dag \hat{c}_i$ plus a fluctuation potential $\hat{V}=\hat{H}-\hat{H}_0$, where $\varepsilon_i$ are the HF orbital energies,  $\hat{c}_i^\dag$ ($\hat{c}_i$) are the creation (annihilation) operators for the electron in the $i$-th HF spin-orbital, respectively, and $\langle ij || kl \rangle$ are the antisymmetrized two-electron repulsion integrals (ERIs) \cite{Szabo:1996}.
 The correlation energy is given by the Rayleigh-Schr\"odinger perturbation series including only the linked terms \cite{Goldstone:57,Shavitt:09},  
\begin{equation}\label{MPn_series}
E_\text{c} = E - E_\text{HF}=\sum_{m=0} \langle \hat{V}(\hat{R}_0 \hat{V})^m\rangle_L,
\end{equation}
 where $\hat{R}_0=(1- |\Phi_0\rangle\langle \Phi_0|)(z-\hat{H}_0)^{-1}$ is the reduced resolvent operator for the HF reference state $|\Phi_0\rangle$  \cite{Shavitt:09,Kutzelnigg:09}. 

\begin{figure}[t]
\centering
    \includegraphics[width =1.06\linewidth, trim = 20 250 0 140]{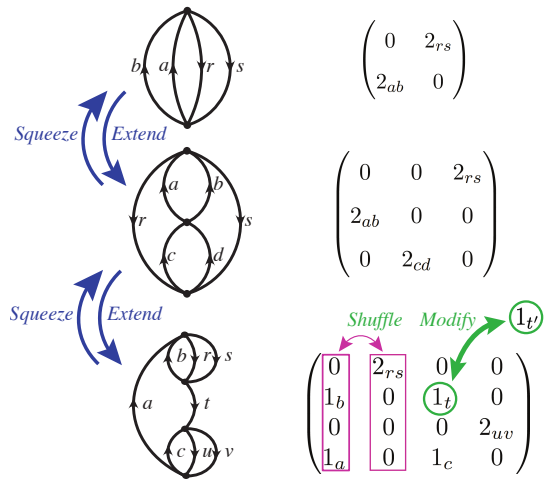} 
\caption{Representative Hugenholtz diagrams contributing to the $n$-order of MP theory with $n=2$, 3, and 4 along with their corresponding adjacency matrices \cite{Tichai:2017,Arthuis:2019} shown to the right of each diagram. The different types of DiagMC updates are indicated by blue arrows (for {\it Extend} and {\it Squeeze}), by magenta arrows (for {\it Shuffle}), and by green arrows (for {\it Modify}), see main text for details.}  
\label{fig:one}
\end{figure}

The different terms in the MP$n$ series of Eq.~(\ref{MPn_series}) are most compactly represented  by means of Hugenholtz diagrams \cite{Negele:1988,Szabo:1996}. Diagrams contributing to the $n$-order consist of $n$ labeled vertices, vertically aligned by convention. Each vertex corresponds to an ERI $\langle ij||kl\rangle$ and has two incoming and two outgoing lines,  corresponding to either particle orbitals $a,b,c,\ldots$ (upward lines) or hole orbitals $r,s,t, \ldots$ (downward lines) \cite{Shavitt:09}.
Additionally, diagrams with different line orientations are considered distinct, a line cannot start and end on the same vertex, each diagram must consist of only one connected component, and its overall sign $s=(-1)^{h- l}$ depends on the number of holes $h$ and of closed loops $l$. Each pair of adjacent vertices contributes the factor $(\sum_h \varepsilon_h - \sum_p \varepsilon_p)^{-1}$, where the sums run over the particles and holes crossing an imaginary horizontal line between the vertices. Finally, each diagram is scaled by $2^{-p}$, $p$ being the number of equivalent line pairs (i.e., co-directed lines that start and end on the same vertex) \cite{Shavitt:09}.  
 The number of $n$-th order Hugenholtz diagrams grows factorially with $n$  \cite{OEIS:A064732}. Representative $n=2{-}4$ diagrams are shown in Fig.~\ref{fig:one}. 
Our DiagMC approach presented below directly samples the expansion of the correlation energy (\ref{MPn_series}) in terms of Hugenholtz diagrams.

We now observe, following recent work on MBPT in nuclear physics \cite{Tichai:2017,Arthuis:2019} that, according to graph theory, Hugenholtz diagrams of order $n$ can be conveniently encoded into $n \times n$ adjacency matrices that satisfy the following conditions: (i) $A_{ij}$ can only take values 0, 1, 2, (ii) $\sum_i A_{ij} = 2 \ \forall  j$, (iii) $\sum_j A_{ij} = 2 \ \forall  i$, and (iv)  $A_{ii} = 0 \ \forall  i$.
Fig.~\ref{fig:one}(b) shows the adjacency matrix representations of selected MP$n$ diagrams. We stress that in some contexts one names `diagram' the summed-over expression, after the sums over the hole and particle indices have been carried out. Here, we call `diagram' an expression depending on these indices, with no sum implied.
We make this apparent by introducing appropriate subscript indices on the entries of the matrices of Fig.~\ref{fig:one}; the diagrammatic rules above, along with the convention we choose for the adjacency matrix, imply that entries below (above) the diagonal will carry hole (particle) indices, respectively. The core idea of the present Letter is to stochastically sample these diagrams -- through their matrix representation -- at all orders, varying the topology of the diagram and the value of the indices \cite{Prokofev:1998,Mishchenko:2000co,Gukelberger:2017en,VanHoucke:2019gm}, converging -- in the statistical sense -- to the exact correlation energy.

For this purpose, we start by relaxing condition (iv) above, considering a larger set of matrices that have $A_{ii}\ne 0$. Within this extended configuration space $\mathcal{E}$, we distinguish between physical matrices -- satisfying all four conditions -- and unphysical ones -- satisfying only conditions (i)-(iii) above. It can be shown \cite{SM} that each matrix in $\mathcal{E}$ can be represented as the sum of two  $N \times N$  permutation matrices $P$  defined by the following conditions (1) Each $P_{ij}=0$ or 1, (2)  $\sum_i P_{ij} = 1 \ \forall  j$, and (3) $\sum_j P_{ij} = 1 \ \forall  i$.
The converse is also, more trivially, true: two permutation matrices always sum to a matrix in $\mathcal{E}$. Therefore, the configuration space $\mathcal{E}$ consists essentially of two copies of the permutation matrix configuration space. As a consequence, we can just design a stochastic process sampling in the permutation matrix configuration space, subsequently `doubling' it to sample over $\mathcal{E}$.

\paragraph{DiagMC procedure.} We now apply the DiagMC methodology \cite{Prokofev:1998,Mishchenko:2000co} by devising a set of updates that can ergodically explore the space of permutation matrices. The \textit{Extend$_\textit{1}$} update adds a row to the bottom and a column to the right of a permutation matrix, thereby going from order $N$ to order $N+1$. We begin by choosing a non-zero entry $P_{ij}$  of the original matrix, setting it to zero, and subsequently `projecting' it onto the newly created column and row. More specifically, we add two new entries $P_{(N+1)j}=1$ and $P_{i(N+1)}=1$. Due to the conventions discussed above, $P_{i(N+1)}$ will carry a hole index, while $P_{(N+1)j}$ will carry a particle index. We then reuse the numerical value of the index of the erased entry $P_{ij}$ as the index carried by one of the two new entries. Depending whether the old value was a particle or hole index, we will need to choose from a discrete uniform random distribution a new hole or particle index, respectively. The probability for this update is then
\beq
\mathcal{P}_{\textit{Extend}_\textit{1}} = \begin{cases} \frac{1}{N \ n_\text{p}} & \mathrm{if} \mbox{ } i \leq j\\
\frac{1}{N \ n_\text{h}} & \mathrm{otherwise}
\end{cases}
\eeq
where $n_\text{h}$ ($n_\text{p}$) is the total number of hole (particle) orbitals in the basis set being used, respectively. For the complementary update, that we denote \textit{Squeeze$_\textit{1}$}, we need to remove the two elements on the last row and column. There is just one way of doing so. Then we need to restore the $P_{ij}$ matrix element whose index might correspond either to a hole or to a particle state, and we can get the numerical value of that index from the index of one of the removed entries. The probability is then, $\mathcal{P}_{\textit{Squeeze}_\textit{1}}=1$.

The \textit{Extend$_\textit{2}$} update adds one column and one row to a permutation matrix, and adds a new `1' entry on the diagonal, on the bottom right. This will always take us to the unphysical sector, and by convention the newly added entry will always carry a hole index. The value is then drawn from a uniform random distribution, and the probability is then $\mathcal{P}_{\textit{Extend}_\textit{2}} = 1 / n_\text{h}$. The complementary \textit{Squeeze$_\textit{2}$} update simply deletes the matrix element in the bottom right corner, returning to an $N \times N$ matrix. There are no probability distributions involved in this process, therefore one  has $\mathcal{P}_{\textit{Squeeze}_\textit{2}} = 1$.

In the \textit{Shuffle} update, we first decide if we want to shuffle rows or columns. We then choose two random rows or column ands swap them. In doing so, the update might need to replace a hole index with a particle one or vice-versa, thus requiring to draw numbers from a uniform distribution. However, since the update is clearly self-complementary, one does not need to keep track of the associated probabilities, since the acceptance ratio depends on weight ratios only.

Lastly, we design a \textit{Modify} update, in which a non-zero hole or particle entry is selected and the associated index is changed to a different value chosen from a uniform distribution. 
This update is also self-complementary \cite{Prokofev:1998,Mishchenko:2000co,Gukelberger:2017en,VanHoucke:2019gm}.

It is easily seen that the set of updates just introduced is ergodic. We then consider two permutation matrices and we apply the updates just introduced to each matrix at each MC step, with the constraint that the two matrices must always have the same dimension. In the spirit of DiagMC, we accept or reject the updates with a probability chosen as to make the process satisfy a detailed balance condition \cite{Prokofev:1998,Mishchenko:2000co,Gukelberger:2017en,VanHoucke:2019gm}; this implies that in the long run the process will spend with each diagram a number of MC steps proportional to the diagram weight, allowing us to collect statistics about the ratio of energies at different orders. The process jumps back and forth between the physical and unphysical sectors, the latter not contributing to the sampled quantities \cite{Gukelberger:2017en,VanHoucke:2019gm}. We verified that at every order the fraction of physical diagrams is always substantially large, moreover an arbitrary unphysical penalty dividing the weight of unphysical diagrams can help in tipping the balance towards the physical sector \cite{Gukelberger:2017en,VanHoucke:2019gm}.

We finally note that there are several distinct ways, in which a given adjacency matrix $\mathcal{A}$ can be represented as a sum of permutation matrices. We will call the number of such ways {\it the multiplicity} of $\mathcal{A}$.
Since the multiplicity is not always one, some diagrams can be incorrectly `counted' more than once. To avoid this spurious multiple-counting we simply divide the weight associated to a matrix by its multiplicity. An algorithmic determination of the multiplicity is presented in the Supplemental Material \cite{SM}.

 \begin{table}[t!]
	\caption{Calculated correlation energies (in $\mu E_h$) for BH and $\mathrm{H}_2\mathrm{O}$ in the 6-31G basis set compared with the reference  MP$n$ data from the Psi4 code.}
	 \begin{tabular}{cccc}
	 \hline
	 \hline
	  Molecule  & MP order $n$  &  DiagMC (this work) &   Exact    \\  \hline 
	BH  &   2  &   $-38.983338  \pm 0.007937$  &  $-38.993128$ \\
	       &   3  &   $-13.297859  \pm 0.004215$  &  $-13.301207$ \\ 
	       &   4  &   $-5.726406  \pm 0.022243 $  & $-5.728702$ \\ 
     	       &   5  &   $-2.664634  \pm 0.050187 $  & $-2.779645$ \\ 
	H$_2$O  &   2  &   $-129.050784  \pm 0.025138$  &  $-129.053394$ \\
	               &   3  &   $-1.550238  \pm 0.010480$  &  $-1.554750$ \\
	               &   4  &   $-5.110730  \pm 0.052355$  &  $-5.247546$ \\
       	       	 \hline
		 \color{black}
	 \end{tabular}
	\label{tab:correlationEnergies}
\end{table}

\paragraph{\textit{Results: correlation energies and scaling.}}

As a first application of the proposed DiagMC/MP$n$ methodology, we carry out proof-of-principle computations on the CH$_2$, H$_2$O, and BH molecules and compare the results with reference MP$n$ calculations to assess the accuracy of the approach.  
This choice of molecules allows us to explore the performance of the DiagMC/MP$n$ approach for different convergence patterns of the MP$n$ series. While CH$_2$ and $\mathrm{B}\mathrm{H}$ are type-A molecules, for which the series converges monotonically, H$_2$O belongs to type~B, exhibiting oscillating convergence \cite{Cremer:96,Leininger:00}.

\begin{figure}[t!]
\centering
\includegraphics[width=1.0\linewidth,trim={0cm 0cm 0cm 0cm},clip]{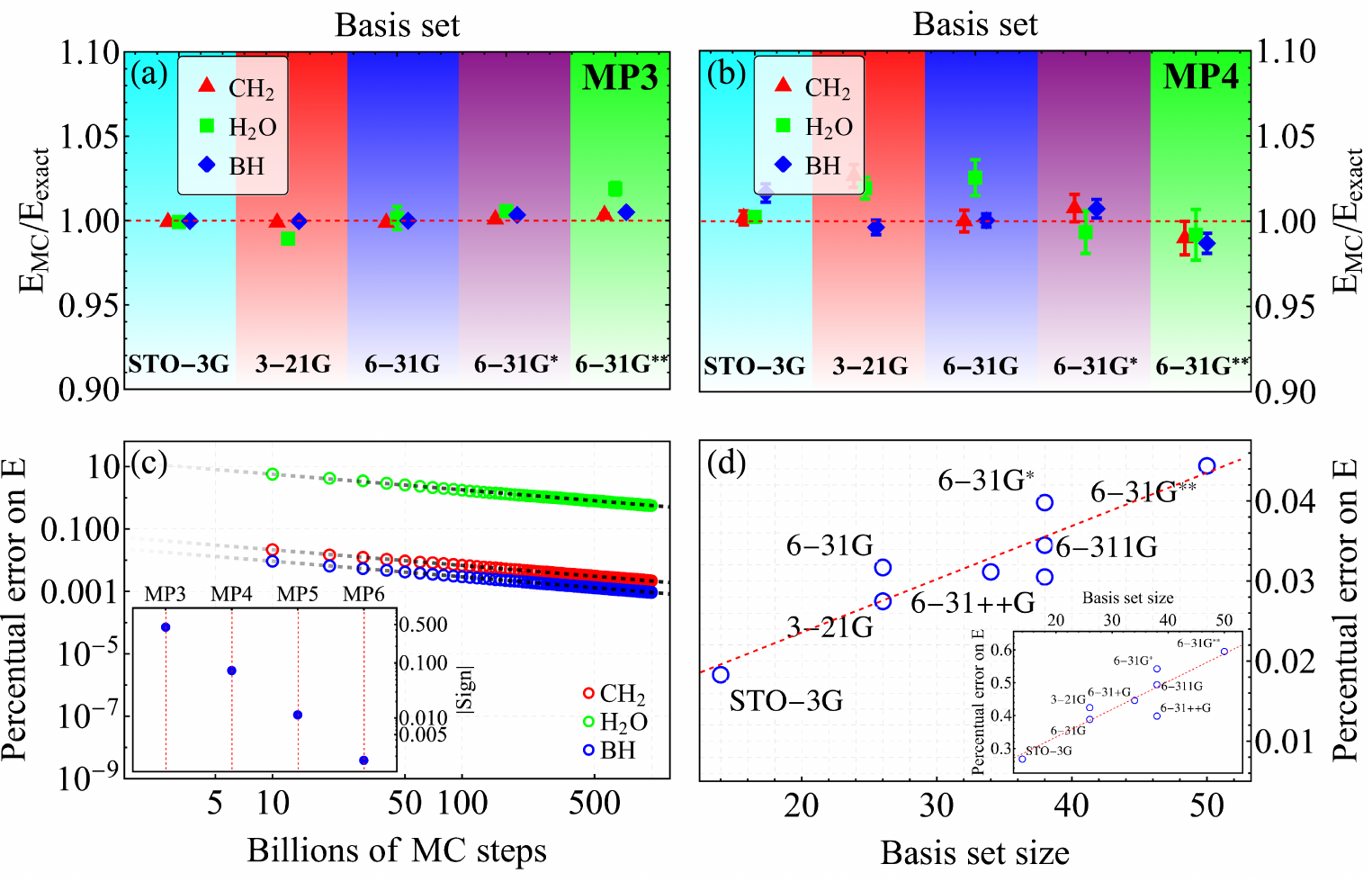}
\caption{(a-b) MP3 and MP4 energies as calculated from our DiagMC procedure for $\mathrm{C}\mathrm{H}_2$, $\mathrm{H}_2\mathrm{O}$ and $\mathrm{B}\mathrm{H}$ molecules and for the STO-3G, 3-21G, 6-31G, 6-31$^{*}$G, 6-31$^{**}$G basis sets, performing $10^{12}$ MC steps per data point. (c) Percentual error on the energy as a function of the number of MC iterations, note the logarithmic scale; the black dashed lines guide the eye and corresponds to statistical $1/\sqrt{N_\text{MC}}$ scaling. Inset: expectation value of the sign as a function of the MP order, notice the logarithmic scale on the vertical axis. (d) Percentual error as a function of the basis size for the $\mathrm{BH}$ molecule after $10^{12}$ MC steps. The red dashed line guides the eyes, highlighting the essentially linear scaling. Main plot shows results at MP3 level, inset at MP4 level.}
\label{fig:two}
\end{figure}

We observe that the DiagMC/MP$n$ correlation energies listed in Table I and shown in Fig.~\ref{fig:two}(a-b), calculated using $10^{12}$ MC steps per data point, are in excellent agreement with the reference MP$n$ data computed using Psi4 \cite{psi4}, which validates all of the elements of our DiagMC procedure described above, regardless of the convergence  pattern of the MP$n$ series. While the error in the DiagMC/MP$n$ correlation energy depends on the molecule and basis set used, it exhibits the expected statistical scaling $1/\sqrt{N_\text{MC}}$ with the number of MC steps,  see Fig.~\ref{fig:two}(c). Figure~\ref{fig:two}(d) and the inset show that, for a fixed $N_\text{MC}$, the error increases approximately  linearly as the basis set size is increased from the smallest (STO-3G) to the largest  (6-31$^{**}$G).
This leads to the overall $O(N^2)$ scaling of the computational effort in our approach with respect to the number of basis states, which makes it much more attractive computationally than conventional MP$n$ ($N^{n+3}$).

\paragraph{Sign problem.} Some of the diagrams we sample have negative weight, therefore we sample with respect to the absolute value of the diagram weight $\mathcal{D}_\xi$ \cite{Gull:2013tg}. 
Doing so, we observe that the statistical error in DiagMC/MP$n$ correlation energies grows significantly with increasing order $n$. This is due to the fermion sign problem, whereby the Hugenholtz diagrams with opposite signs cancel out, making it necessary to use an increasingly large number of MC steps to obtain a nonzero signal-to-noise ratio \cite{Landau:00,Loh:90,Chandrasekharan:99}. For instance, the contributions to the MP6 energy for the NH molecule using the 6-31G basis set and $10^9$ MC steps are estimated to be $0.342020 \pm 0.001172$ and $0.342260 \pm 0.001262$, respectively, with a sensibly increased error when the two values are subtracted to calculate the actual MP6 energy \footnote{All the errors on MC quantities we report throughout this Letter have been determined by means of a jackknife analysis, using the ALEA library from the ALPSCore package \cite{Bauer_2011,Gaenko:2017uf,Wallerberger:2018uf}.}. We have also investigated this analytically, verifying that several topologies are dominated by near-perfect cancellations.

This phenonomenon, bearing a remarkable resemblance with the sign problem observed in other contexts \cite{Prokofev:2008it,Kozik:2010fl,VanHoucke:2010kya,VanHoucke:2012ica,Vlietinck:2014jf,VanHoucke:2019gm} is, however, distinct from -- and less severe than -- the one that plagues quantum many-body MC simulations of, e.g., Fermi-Hubbard models. There, one is interested in the thermodynamic limit, and the expectation value of the sign decreases exponentially with the size of the system \cite{Landau:00,Chandrasekharan:99,Loh:90}. In contrast, for finite-size molecules explored here, this expectation value is small, decreases exponentially with the perturbation theory order, but is always finite, as shown in the inset of Fig.~\ref{fig:two}(c), significantly reducing the acuity of the sign problem. More in detail, here the sign problem is completely absent in MP2, quite moderate in MP3, and largely caused by the singles and triples contributions in MP4, see \cite{SM} for a detailed analysis. There we also analyze how the the main idea behind the CDet algorithm \cite{Rossi:2017fd} could lead to a substantial mitigation of the sign problem in the present context.

\paragraph{Outlook and conclusions.} 
We have demonstrated a low-scaling stochastic approach to calculating molecular electronic correlation energies based on DiagMC sampling of the MP$n$ series. The approach samples the many-body electronic correlation energy directly using Hugenholtz diagrams, encoded in adjacency matrices using combinatorial graph theory \cite{Arthuis:2019}. Our DiagMC/MP$n$ approach shares many of the attractive features with its antecedents in quantum many-body physics  \cite{Prokofev:1998,VanHoucke:2010kya}, such as low scaling and the ability to converge towards the exact result (the full CI limit). We demonstrate accurate results for the MP$n$ correlation energies with $n \lesssim 5$.  Already at MP4 level, the accuracy of our results is comparable to those provided by the CCSD(T) approach (the ``golden standard'' of quantum chemistry \cite{Bartlett:07,Hesselmann:19}). Thus, our low-scaling DiagMC/MP$n$ methodology could be applied to a wide range of quantum chemical problems, where high-precision estimates of dynamical correlation energy are crucial, such as calculating intermolecular dispersion interactions \cite{Hesselmann:19}. 

We find that results for $n \gtrsim 5$ are affected by the sign problem, which, however, is significantly less severe than the sign problem encountered in the thermodynamic limit {\cite{Loh:90,Landau:00}} due to the finite size of  molecular systems. In future work, we plan to address this problem adapting the recently-developed CDet algorithm {\cite{Rossi:2017fd,Rossi:2018uza}}. This would enable one to perform reliable extrapolations to the full CI limit \cite{Cremer:96}, using, e.g., Pad\'e approximants, resummation techniques, and Feenberg scaling \cite{Cremer:11}, and to explore the convergence behavior of the MP$n$ series for large molecules, currently outside of reach of modern quantum chemistry techniques. Thus far, the behavior of the MP$n$ series for large $n$ has been explored only for the smallest molecules, for which full CI calculations could be performed {\cite{Leininger:00}}.



\begin{acknowledgments}
\paragraph{Acknowledgements.} We acknowledge stimulating discussions with Sergey Varganov, Artur Izmaylov, Jacek K{\l}os,  Piotr \.Zuchowski, Dominika Zgid, Nikolay Prokof'ev, Boris Svistunov, Robert Parrish, and  Andreas He{\ss}elmann at various stages of this work.
G.B.~acknowledges support from the Austrian Science Fund (FWF), under project No.~M2641-N27.
Q.P.H.~acknowledges support from the Austrian Science Fund (FWF), under project No.~M2751.
M.L.~acknowledges support by the Austrian Science Fund (FWF), under project No.~P29902-N27, and by the European Research Council (ERC) Starting Grant No.~801770 (ANGULON).
This work is supported by the Deutsche Forschungsgemeinschaft (DFG, German Research Foundation) under Germany's Excellence Strategy EXC2181/1-390900948 (the Heidelberg STRUCTURES Excellence Cluster).
The authors acknowledge support by the state of Baden-W\"urttemberg through bwHPC.
\end{acknowledgments}


%

\end{document}


\widetext
\begin{center}
\textbf{\large Supplemental Material:\\[5pt] Diagrammatic Monte Carlo for electronic correlation in molecules: \\ high-order many-body perturbation theory with low scaling}
\end{center}
\setcounter{equation}{0}
\setcounter{figure}{0}
\setcounter{table}{0}
\setcounter{page}{1}
\makeatletter
\renewcommand{\theequation}{S\arabic{equation}}
\renewcommand{\thefigure}{S\arabic{figure}}
\renewcommand{\bibnumfmt}[1]{[S#1]}
\renewcommand{\citenumfont}[1]{S#1}
\makeatother

\section{Adjacency matrices and permutation matrices}

Here, we demonstrate the claim made in the main text, that each adjacency matrix corresponding to a Hugenholtz diagram can be written as the sum of two permutation matrices.

\begin{defn}
A square matrix is said to be a permutation matrix if its entries are either 1 or 0, and moreover, the sum of each row/column is exactly 1.
\end{defn}

A adjacency matrix in the extended configuration space $\mathcal{E}$ defined in the main text is a 2-permutation matrix in the following definition.

\begin{defn}
A square matrix is said to be a 2-permutation matrix if its entries are either 2, 1, or 0, and the sum of each row/column is exactly 2.
\end{defn}

Clearly, the sum of any two permutation matrices is a 2-permutation matrix. We will call such 2-permutation matrices decomposable. Our goal is to show that in fact, all 2-permutation matrices are decomposable.

\begin{prop} \label{prop:main_prop_decomposition}
Any 2-permutation matrix is decomposable, i.e.~it can be expressed as a sum of two permutation matrices.
\end{prop}

We will use Lemmas~\ref{lem:deleting_2} and \ref{lem:invariance_under_permutations} below, which are straightforward.

\begin{lem}\label{lem:deleting_2}
Let $M$ be a 2-permutation matrix and $M'$ obtained from $M$ by deleting all rows and columns containing an entry of value 2. Then $M$ is decomposable if and only if $M'$ is. 
\end{lem}

This lemma allows us to reduce to the case where $M$ only contains 1's. We will assume this from now on.

\begin{lem} \label{lem:invariance_under_permutations}
Let $M$ be a 2-permutation matrix and $M'$ obtained from $M$ by a series of row/column permutations. Then $M$ is decomposable if and only if $M'$ is.
\end{lem}

\noindent\paragraph{\textbf{Proof of Proposition~\ref{prop:main_prop_decomposition}.}} Let $M$ be a 2-permutation matrix. We will show that $M$ is decomposable. As mentioned above, we can, and we will, assume that $M$ only contains 0's and 1's. By Lemma~\ref{lem:invariance_under_permutations}, it suffices to do so after a series of row and column permutations. More precisely, by doing row/column permutations, we will turn $M$ into a new matrix with 1's on the diagonal. The decomposition of such a matrix into two permutation matrices is obvious.

Starting with the first row, by doing a row/column permutation, we can turn $M$ into a new matrix such that the top left is given by

\[
\begin{bmatrix}
1 \\
1
\end{bmatrix}
\]
Necessarily, anything else on the first column besides the first two entries are 0's like so
\[
\begin{bmatrix}
1 \\
1 \\
0 \\
0 \\
\vdots \\
0
\end{bmatrix}
\]

Now, for the second row, we know that there exists a non-zero entry. So by row/column permutation, we can turn $M$ into a new matrix with the first two columns having one of the following two forms

\[
\begin{bmatrix}
1 & 1\\
1 & 1 \\
0 & 0 \\
0 & 0 \\
\vdots & \vdots \\
0 & 0 \\
\end{bmatrix} \text{  or  } \begin{bmatrix}
1 & 0\\
1 & 1 \\
0 & 1 \\
0 & 0 \\
\vdots & \vdots \\
0 & 0 \\
\end{bmatrix}
\]

Now it is clear how to proceed inductively. The main point is that suppose we are done with the first $k$-rows, the first $k$ entries of the $k+1$-th row will have one of the following two forms

\begin{equation}
\begin{bmatrix}
0 & 0 & \cdots & 0 & 1
\end{bmatrix} \text{  or  } \begin{bmatrix}
0 & 0 & \cdots & 0 & 0
\end{bmatrix} \label{eq:2_form_for_row_k+1}
\end{equation}
i.e. there is at most one 1 entry here, which means that we can find another 1 on the same row. Let $(k+1,l)$ denote the coordinates of that entry. Note that $l \geq k+1$.

The $l$-th column will have another 1 at coordinate $(k',l)$. If $k'<k+1$, then we use column permutation to swap the $l$-th column and the $(k+1)$-th column. If $k'>k+1$, then we use row permutation to swap the $k'$-th row and the $(k+2)$-th row, and then, use column permutation to swap the $l$-th column and the $(k+1)$-th column. After doing this we see that the $(k+2)$-th row will have the form \eqref{eq:2_form_for_row_k+1} again, where, of course, in the first case, the 1 is on the $(k+1)$-slot. Now, we can continue the process until we are done. In the end, our matrix will have the following block(s) along the diagonal (and $0$'s everywhere else)
\[
B = \begin{bmatrix}
1 & 0 & 0 & \cdots & 0 & 1 \\
1 & 1 & 0 & \cdots & 0 & 0 \\
0 & 1 & 1 & \cdots & 0 & 0 \\
\vdots & \vdots & \vdots & \ddots & \vdots & \vdots \\
0 & 0 & 0 & \cdots & 1 & 0 \\
0 & 0 & 0 & \cdots & 1 & 1
\end{bmatrix}.
\]
It is clear that this matrix is decomposable and the proof concludes.
\hfill$\square$

\section{Counting multiplicities}

Above, we showed that a 2-permutation matrix $A$ could be written as a sum of two permutation matrices. This was done by showing that by permuting rows and columns of $A$, we can bring it to a special form, from which the decomposition of $A$ into a sum of two permutation matrices could be obtained directly. To determine the multiplicity of a 2-permutation matrix, i.e.~to determine the number of ways to do that, we proceed similarly. Namely, since permuting rows and columns does not change the result, we can count the number of decompositions after $A$ has been transformed to a special form.

In what follows, we will assume that $A$ does not have $2$-entries, as those are easy to deal with. From the proof of the previous Section, we see that by permuting rows and columns, any $2$-permutation matrix $A$ could be brought to block matrix, where the diagonal is given by blocks of the form
\[
B = \begin{bmatrix}
1 & 0 & 0 & \cdots & 0 & 1 \\
1 & 1 & 0 & \cdots & 0 & 0 \\
0 & 1 & 1 & \cdots & 0 & 0 \\
\vdots & \vdots & \vdots & \ddots & \vdots & \vdots \\
0 & 0 & 0 & \cdots & 1 & 0 \\
0 & 0 & 0 & \cdots & 1 & 1
\end{bmatrix}
\]
It thus remains to count the number of matrices like $B$ could be written as a sum of two permutation matrices. The final answer for $A$ will be the product of all those numbers for all the blocks.

Suppose we want to write $B=B' + B''$, where $B'$ and $B''$ are permutation matrices. We will now build $B'$ column by column. For the first column, there are exactly two choices for $B'$, corresponding to having $1$ on the first or second row.

\begin{enumerate}
\item Suppose we choose the $1$ on the first row, we see that for the second column, we have exactly one choice: picking $1$ on the second row (since otherwise, $B'$ will never have a $1$ on the second row). Continuing this way, we see that there is exactly one way to choose the rest of $B'$.

\item Suppose we choose the $1$ on the second row, arguing similarly, we also see that the rest of $B'$ is determined uniquely. Thus, there is exactly one way to get the rest of $B'$.
\end{enumerate}

Altogether, we see that there are exactly two ways to write $B$ as a sum of two permutation matrices. Thus, the multiplicity of $A$ is $2^n$, where $n$ is the number of blocks.

\section{Sign problem}

In this Section, we examine the sign problem in Diag/MP$n$ simulations and suggest a promising path towards its resolution building on recent advances in connected determinant DiagMC (CDet) \cite{Rossi:2017fd}. As already stressed in the main text, we note from the outset that the sign problem in Diag/MP$n$ is different from that encountered in previous fermionic DiagMC simulations (see, e.g., Ref.~\cite{Rossi:2017fd}). The latter are typically performed at finite temperature and in the thermodynamic limit, by sampling imaginary-time Feynman diagrams. In contrast, here we are interested in systems with a finite number of electrons at zero temperature described by the Hugenholtz diagrams. 

The sign problem here manifests itself as a rapid (exponential or factorial) decline, as a function of the MP order $n$, of the average value of the sign of sampled diagrams, caused by a nearly perfect cancellation between the positive and negative diagrams \cite{Loh:90,Rossi:2017fd}. This decline causes a rapid increase in statistical error, if the number of MC steps is kept fixed, for the quantity being calculated (in our case, the correlation energy) making DiagMC/MP$n$ simulations for $n \gtrsim 6$ very computationally demanding. The inset of Fig.~2(c) in the main text shows the average value of the sign as a function of MP$n$  order. The sign problem becomes quite relevant at the 4th order, hence we will focus on MP4 diagrams in the following.

To gain more insight into how sign cancellations occur between the individual Hugenholtz diagrams, we define the quantity 
\begin{equation}\label{Rs}
R_{\mathcal{S}} = \frac{ | \sum_{i\in \mathcal{S}} \mathcal{D}_i | }{ \sum_{i\in \mathcal{S}}  |\mathcal{D}_i| }
\end{equation}
where $\mathcal{S}$ denotes a subset of all diagrams and $\mathcal{D}_i$ is the diagram weight associated to the diagram $i$, as discussed in the main text. From Eq.~(\ref{Rs}) one immediately sees that $R_{\mathcal{S}}$ will be close to unity if no cancellations occur between the diagrams in the subset $\mathcal{S}$.  In contrast, $R_{\mathcal{S}}$ will approach zero if the sum of the diagram in $\mathcal{S}$ has many cancellations, and will be exactly zero in the case of perfect cancellation.
We note that these quantities are not additive, i.e., 
given two subsets of all diagrams, be them $\mathcal{S}_A$ and $\mathcal{S}_B$, the conditions $R_{\mathcal{S}_A} \sim 1$ and $R_{\mathcal{S}_B} \sim 1$ do not necessarily imply $R_{\mathcal{S}_A\cup \mathcal{S}_B} \sim 1$, since cancellations may occur between the diagrams when the subsets $\mathcal{S}_A$ and $\mathcal{S}_B$ are joined together.
 
We now want to Eq. (\ref{Rs}) to estimate the degree of cancellation between the different subsets of Hugenholtz diagrams, in order to propose a path towards a solution. Preliminarly, we assign a progressive integer number -- let us call it the diagram ID -- to each order 4 diagram topology in the following way: as discussed in the main text, the adjacency matrix of each diagram is the sum of two permutation matrices. At order 4 there are $4!=24$ different permutation matrices, and after decomposing a diagram in two permutation matrices, we assign to each one an index -- be them $i$ and $j$ -- according to the canonical ordering of the permutation it represents. Finally we assign to each diagram topology the integer number $24 i + j$. It turns out that different numbers can correspond to the same adjacency matrix, see Section II of the present Supplemental Material on the multiplicity: in this case, we conventionally assign to a diagram topology the smallest integer, which will be the diagram ID. This establishes a 1-to-1 mapping between a subset of integers and the diagram topologies at order 4.

\begin{figure}[t]
\centering
\includegraphics[width=0.4\linewidth]{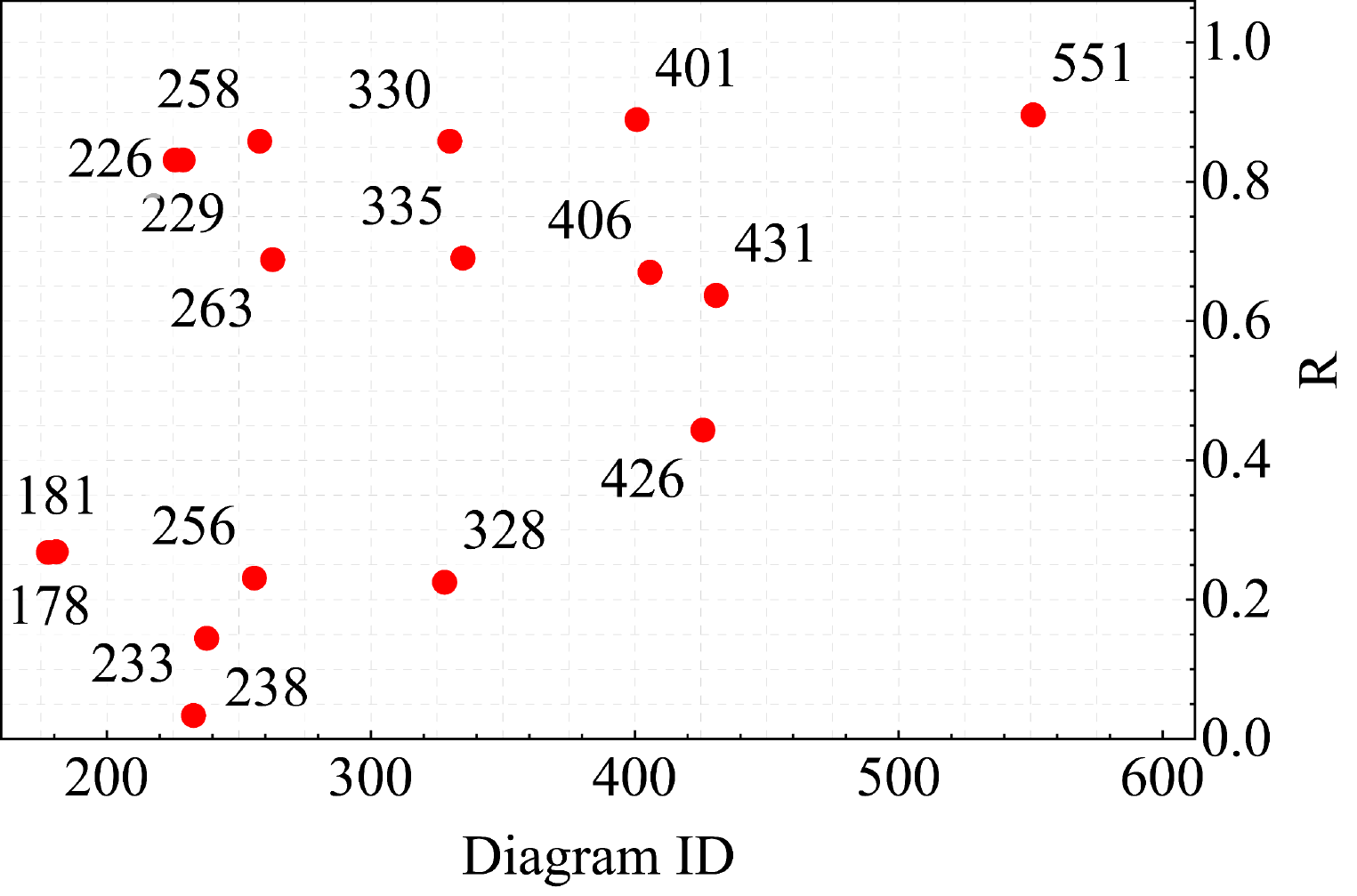}
\hspace{36pt}
\includegraphics[width=0.4\linewidth]{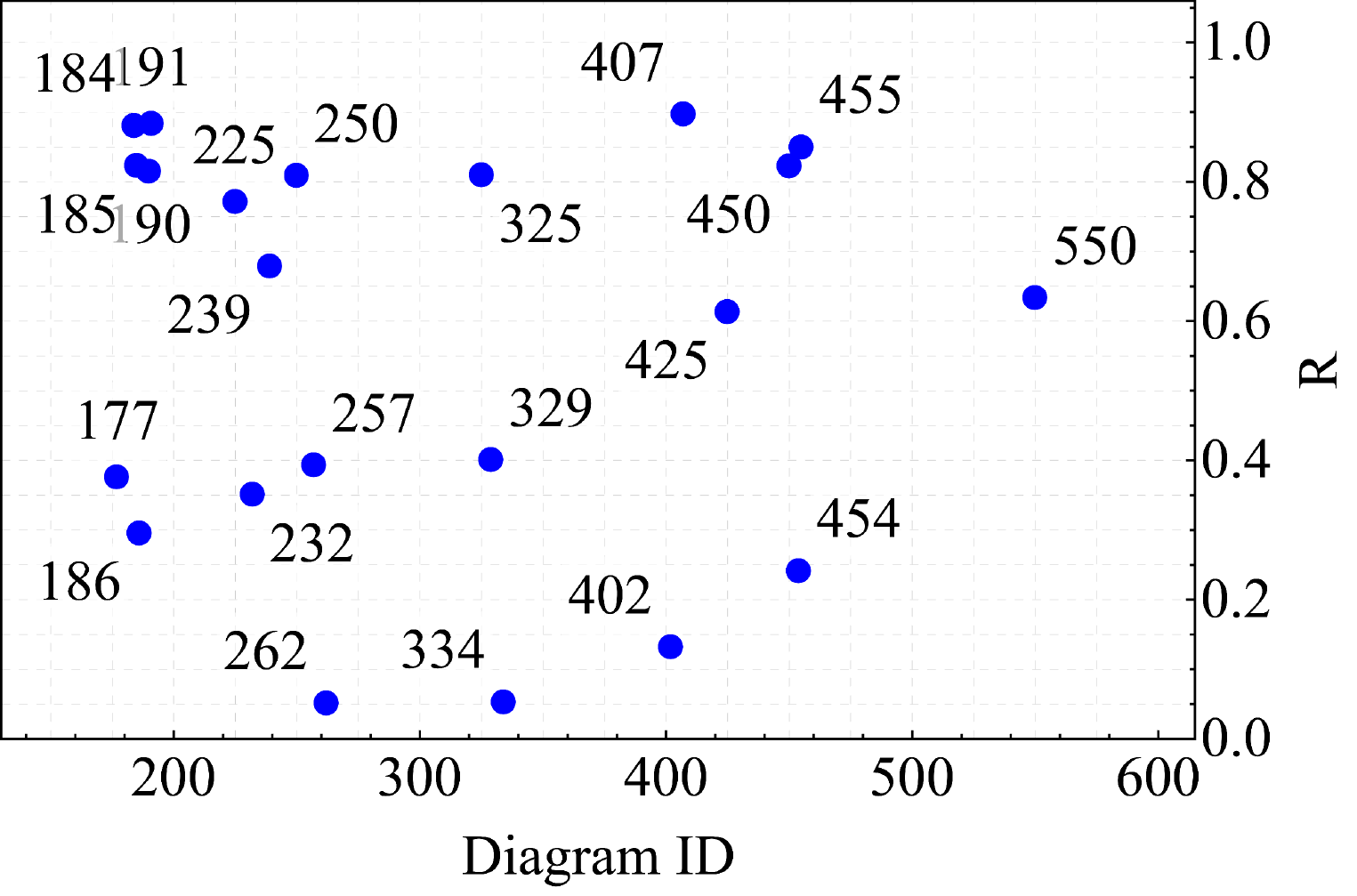}
\caption{The $R$ factors defined by Eq. (\ref{Rs}) plotted vs. diagram IDs for MP4 Hugenholtz diagrams summed over their internal indices for the BH molecule in the 6-31G basis. Diagrams topologies with total positive expectation value of the sign are plotted on left panel, diagrams topologies with total negative expectation value of the sign on the left.} 
\label{fig:R_scatter_plot}
\end{figure}

Having done so, in Fig.~S1 we show a scatter plot of the $R$ values for the distinct 39 fourth-order Hugenholtz diagram topologies, as a function of their diagram ID, with the subsets defined by all the diagrams with a fixed topology but different internal indices. We observe that the vast majority of the diagrams have $R$ values above 0.2. Notable exceptions include diagram topologies with ID 233, 262 and 334, which correspond to the `single' diagram 4, and to the `triple' diagrams 23 and 24 in Fig.~5.6 of Ref.~\cite{Shavitt:09}.  These diagram topologies have $R \lesssim 0.05$, signaling a large number of cancellations between positive and negative contributions, in turn meaning that the stochastic sum over the internal variables will be affected by a considerably large statistical error. Also, these diagram topologies contribute the most to the statistical error in the MP4 correlation energy, since the final error is dominated by the terms with the largest error.

This suggests that one could greatly mitigate the sign problem by identifying `problematic' topologies and replacing, for these topologies, the stochastic sums with actual sums. Extending this idea, one could imagine a mixed approach where the stochastic process moves through different topologies, whereas all sums over the internal indices are carried over exactly. Using the notation of Ref.~\cite{Rossi:2017fd}, denoting with $\mathcal{T}$ the topology of a diagram, and with $\mathcal{X}$ all internal variables, this would correspond to sampling configurations $\mathcal{C} = (\mathcal{T})$ according to the distribution
\beq
\mathcal{P} (\mathcal{C}) = |\sum_{\mathcal{X}} \mathcal{D(\mathcal{T}; \mathcal{X})}| \; ,
\eeq
motivated by cancellations between diagrams with the same topology, whereas standard diagrammatic Monte Carlo samples configurations $\mathcal{C} = (\mathcal{T}, \mathcal{X})$ according to the distribution
\beq
\mathcal{P} (\mathcal{C}) = |\mathcal{D(\mathcal{T}; \mathcal{X})}| \; .
\eeq
It is interesting noting that this approach would be complementary to the CDet approach \cite{Rossi:2017vk}, which samples configurations $ \mathcal{C} = (\mathcal{X}) $ according to the distribution
\beq
\mathcal{P} (\mathcal{C}) = |\sum_{\mathcal{T}} \mathcal{D(\mathcal{T}; \mathcal{X})}| \; ,
\eeq
giving rise to cancellation between diagrams with \textit{different} topologies. The CDet strategy could be adapted to the present case, as well.

In conclusion, it is worth mentioning that we find that significant cancellations occuring when the diagrams with different topologies are added together. For example, while the individual $R_{\mathrm{S}_i}$ values for the singles diagram topologies S$_1$-S$_4$ as defined in Fig.~5.6 of Ref.~\cite{Shavitt:09} are all above 0.5, but we have $R_{\mathrm{S}_1\cup \ldots\cup {\mathrm{S}_4}}=0.016$. This motivates our ongoing studies aimed at further characterizing which diagrams give rise to most of the cancellations in the MPn series, order by order, with potential far-reaching application both for analytical and stochastical techniques.


%